\documentclass{aastex61}
\usepackage{multirow}
\usepackage{captcont}
\received{XXX}
\revised{XXX}
\accepted{XXX}
\submitjournal{ApJ}

\shorttitle{CO gaps}
\shortauthors{van der Marel et al.}

\begin{document}

\title{Rings and gaps in protoplanetary disks: planets or snowlines?}

\correspondingauthor{Nienke van der Marel}
\email{astro@nienkevandermarel.com}

\author{Nienke van der Marel}
\affil{Herzberg Astronomy \& Astrophysics Programs, 
National Research Council of Canada, 
5071 West Saanich Road, 
Victoria BC V9E 2E7,
Canada}

\author{Jonathan P. Williams}
\affil{Institute for Astronomy, 
University of Hawaii, 
2680 Woodlawn dr.,
96822 Honolulu HI,
USA}
\nocollaboration

\author{Simon Bruderer}
\affil{Max-Planck Institut f{\"u}r Extraterrestrische Physik, 
Giessenbachstrasse 2, 
85741 Garching bei M{\"u}nchen,
Germany}
\nocollaboration

\begin{abstract}
High resolution ALMA observations of protoplanetary disks have revealed that many, if not all primordial disks consist of ring-like dust structures. The origin of these dust rings remains unclear, but a common explanation is the presence of planetary companions that have cleared gaps along their orbit and trapped the dust at the gap edge. A signature of this scenario is a decrease of gas density inside these gaps. In recent work, \citet{Isella2016} derived drops in gas density consistent with Saturn-mass planets inside the gaps in the HD~163296 disk through spatially resolved CO isotopologue observations. However, as CO abundance and temperature depends on a large range of factors, the interpretation of CO emission is non-trivial. We use the physical-chemical code DALI to show that the gas temperature increases inside dust density gaps, implying that any gaps in the gas, if present, would have to be much deeper, consistent with planet masses $>$M$_{\rm Jup}$. Furthermore, we show that a model with increased grain growth at certain radii, as expected at a snowline, can reproduce the dust rings in HD~163296 equally well without the need for companions. This scenario can explain both younger and older disks with observed gaps, as gaps have been seen in systems as young $<$1 Myr. While the origin of the rings in HD~163296 remains unclear, these modeling results demonstrate that care has to be taken when interpreting CO emission in protoplanetary disk observations.
\end{abstract}

\keywords{Astrochemistry - Protoplanetary disks - Stars: formation - ISM: molecules}


\section{Introduction}
In the last few years, ALMA has revolutionized the field of study of protoplanetary disks, the birth cradles of planets. High-resolution imaging has revealed that the continuum emission of the millimeter-dust grains is not smooth, but consists of multiple dust rings. The most well-known example is the young HL~Tau disk \citep{HLTau2015}, the first disk resolved at 25 mas resolution at millimeter wavelengths. Other disks showing multiple rings are e.g. TW Hya \citep{Andrews2016}, HD~163296 \citep{Isella2016}, HD~169142 \citep{Fedele2017}, AA~Tau \citep{Loomis2017} and AS~209 \citep{Fedele2018}. Even in lower resolution data ($\sim$0.2"), evidence for outer dust rings is often found, e.g. in HD~100546 \citep{Walsh2014}, RXJ~1615-3255 \citep{vanderMarel2015-12co} and HD~97048 \citep{vanderPlas2017}. 

Whereas these rings and gaps are clearly common, the mechanism to explain the origin of the gaps remains unclear. Proposed scenarios include dust growth in condensation zones or snowlines \citep{Zhang2015}, zonal flows \citep{Flock2015}, self-induced dust pile-ups \citep{Gonzalez2017}, aggregate sintering \citep{Okuzumi2016}, large scale instabilities \citep{LorenAguilar2016} or secular gravitational instabilities \citep{Takahashi2016} and, most popular, clearing by embedded planets \citep[e.g.][]{LinPapaloizou1979,KleyNelson2012,Baruteau2014}, including triggering of multiple gaps by a single planet \citep{Dong2017multi,Bae2017}. However, detections of planets embedded in disks are rare \citep[e.g.][]{KrausIreland2012,Quanz2013,Currie2015,Sallum2015} and often only upper limits can be set \citep{Testi2015,Maire2017,Pohl2017}. On the other hand, giant planets at wide orbits are rare \citep{Bowler2016} and the observed structures may be caused by currently undetectable, low-mass planets \citep{Dong2015gaps,Rosotti2016,Dipierro2017,DongFung2017}. Furthermore, the commonality of ring structures at the very early and very late stages of protoplanetary disks ($<$1 to $>$ 10 Myr) casts doubts on the explanation of planets, as planets would need to be formed in $\ll$1 Myr time scales, well below current predictions of planet formation theory \citep{Helled2014}.

In order to understand the origin of the gaps, the structure of the gas needs to be known: whereas planets are expected to lower the gas density inside the gaps, other effects such as snow lines would not change the density itself. Gas cavities (linked to giant planets) have been revealed through CO observations in transition disks with large dust cavities \citep[e.g.][]{vanderMarel2016-isot,Dong2017,Fedele2017}, but the gas structure inside these narrow gaps has remained difficult to constrain. Gas gaps have been claimed through CO 2--1 isotopologue data in HD~163296 \citep{Isella2016}. In their modeling procedure, both the CO abundances and gas temperatures were parametrized. The depth of these gaps can be linked directly to planet masses of embedded planets \citep{Fung2014,Rosotti2016,DongFung2017}. \citet{Isella2016} deduced that $\sim$Saturn mass planets must be responsible for the gap structure, based on the depth of the gas gaps. However, kinematic signatures point towards more massive, Jupiter-like planets \citep{Teague2018} inside the gaps and direct imaging with Keck/NIRC2 has revealed a point source at 0.5" consistent with a Jupiter-like planet as well \citep{Guidi2018}.

Clearly, \emph{gaps in CO emission cannot be converted directly into gas surface density drops}: the CO abundance is not constant with respect to H$_2$ throughout the disk due to e.g. photodissociation, freeze-out and chemical effects, and the gas temperature is decoupled from the dust temperature in the bulk of the disk depending on the local conditions, due to various heating-cooling effects \citep{Zadelhoff2001,Aikawa2002}. The gas temperature can be up to two orders of magnitude higher in the disk atmosphere \citep{Bruderer2012,Bruderer2013}. Recently, \citet{Facchini2017gaps} showed that the gas temperature is in fact \emph{lower} inside gas gaps created by giant planets due to the low dust-to-gas ratio and consequently, decreased gas-dust energy transfer, using the DALI code including dust evolution \citep{Bruderer2013,Facchini2017rout}. This implies that care should be taken when drops in CO emission are observed. A physical-chemical model is thus crucial to interpret the complex structure CO emission properly.

In this letter, we present the modeling outcome of two mechanisms for a dust ring disk, using the DALI code with a parametrized density structure. We show how the CO emission changes due to the presence of dust gaps on one hand and due to changing in the grain-size distribution on the other hand. We discuss the implications for constraining the origin of dust rings from high resolution CO observations. 

\section{Model}
The DALI code \citep{Bruderer2012,Bruderer2013} was developed for the prediction of molecular line fluxes of protoplanetary disks, for a given surface density structure of gas and dust and a given stellar spectrum providing the radiation field. DALI solves for the dust radiative transfer, the chemical abundance, the molecular excitation and the thermal balance to obtain the gas temperature and CO abundances at each position in the disk. Further details on the parameters and assumptions in DALI are given in \citet{Bruderer2013}.

\begin{figure}[!ht]
\begin{center}
\includegraphics[width=\textwidth]{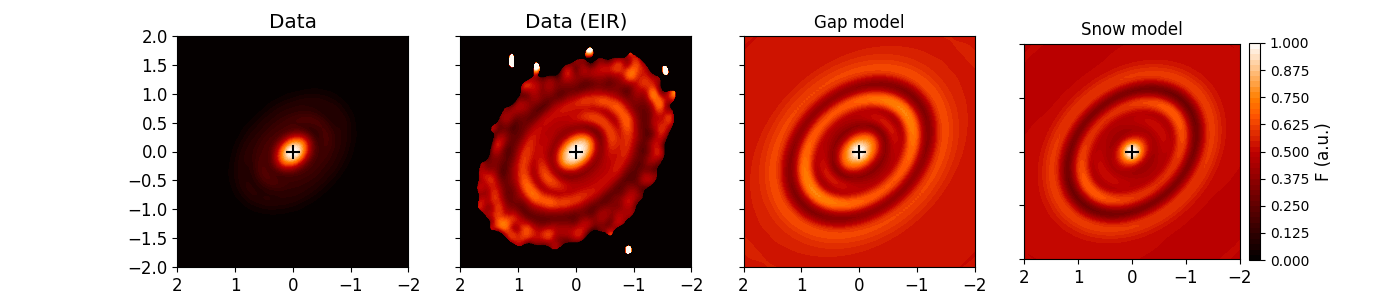}
\caption{Dust continuum flux density (230 GHz) of the data in the original (first) and enhanced image representation/EIR (second), Model A (gap model) and Model B (snow model), convolved to the beam of the observations and in the EIR as described in the text. The morphology of the dust rings is reproduced in both models.}
\label{fig:dustmodels}
\end{center}
\end{figure}

For our model, we mimic the structure of the HD~163296 disk as analyzed by \citet{Isella2016}. The most recent parallax of HD~163296 is 9.85$\pm$0.11 mas using \emph{Gaia} DR2, corresponding to a distance of 101.5$\pm1.2$ pc \citep{Gaia2018}. As the old distance was 122 pc  \citep{vandenAncker1997}, the gap radii and the stellar luminosity as derived by \citet{Isella2016} are scaled accordingly. We do not aim to provide a perfect fit to the observations, but use the structure to illustrate how the CO emission changes due to changes in the dust structure (left two panels of Figure \ref{fig:dustmodels}). We use the same data images as presented in \citet{Isella2016} for our comparison. In order to enhance the visibility of the ring features, for each image we compute the ratio between the original and a smoothed version (smoothing FWHM is twice the beam size of the observations) and normalize the result. Both these enhanced image representation (EIR) and original images are provided to compare with the models. The central deficit in the $^{13}$CO and C$^{18}$O images is not a real gap in the CO, but is caused by the continuum subtraction (see discussion in \citet{Boehler2017}). 

For the model, we assume a surface density profile $\Sigma(r)$ as a radial power-law with exponential cut-off following \citet{LyndenPringle1974}:
\begin{equation}
\Sigma(r) = \Sigma_c \left(\frac{r}{r_c}\right)^{-\gamma} {\rm exp}\left(-\left(\frac{r}{r_c}\right)^{2-\gamma}\right)
\end{equation}
with $\gamma=1$. The dust surface density is computed by dividing the equation by the gas-to-dust ratio. Settling is parametrized following \citet{Bruderer2013}, with the scale height of the large grains distributed at a fraction of the total scale height. Furthermore, we assume the dust gap radial locations as derived in \citet{Isella2016} scaled to the new \emph{Gaia} distance of 101.5 pc. The assumed parameters and properties of HD~163296 are summarized in Table \ref{tbl:params} and the observational properties in Table \ref{tbl:obs}. Stellar parameters are taken from \citet{Mendigutia2013}.

The model is set up in two different ways: in Model A ('gap model') the dust surface density is decreased by a factor 0.010, 0.014 and 0.17 for gap $r_1$, $r_2$ and $r_3$ respectively, whereas the gas surface density is decreased by a factor 0.4, 0.29 and 0.56 respectively, following the dust and gas model of \citet{Isella2016}. In Model B ('snowline model'), the dust surface density remains the same, but the fraction of large grains in the midplane $f_{ls}$ is set to 0.01 inside the three gaps, and otherwise $f_{ls}$=0.99, to mimic increased dust growth in the rings and a deficit of large grains in the gaps. The two dust populations used in DALI are the small grain population (0.005 - 1.0$\mu$m) and large grain population (0.005 - 1000$\mu$m), and $f_{ls}$ corresponds to the mass fraction of the large grains w.r.t. to the total mass of dust grains. Note that in Model A $f_{ls}$ was set constant at 0.85. This model mimics the dust size distribution as calculated from dust evolution models by \citet{Pinilla2017-ice}, where the large grain fraction is higher at certain radii as the result of snow lines and the resulting change in fragmentation velocity. The radii are chosen based on the observed ring radii rather than the actual temperature profile at this point.

Snowlines are expected to induce pressure bumps in viscosity gradients, due to the change in the gas ionization fraction as a result of the change in chemistry \citep[e.g.][]{Flock2015}, which result in dust traps as well. As the calculation of the efficiency of this effect is beyond the capabilities of DALI, no additional change in gas or dust surface density is introduced in Model B. The choice of $f_{ls}$=0.01 is justified by testing a range of models with different values for $f_{ls}$: a value of 0.01 or lower is required to reproduce the dust continuum contrast seen in the rings of the observations. Values of 10$^{-2}$--10$^{-4}$ have been tested and the difference in CO emission is negligible. In addition, we have run Model 0 without any radial changes in the surface density or grain size distribution, and a variation of Model A with much deeper gaps (Model C). Figure \ref{fig:DALIplots} shows the structure in each model. 

DALI was run using a grid with 190 cells in radial and 60 cells in vertical direction, using time-dependent chemistry up to 1 Myr. The additional DALI modules with isotope-selective photodissociation \citep{Miotello2014} were not included in all models, but this is not expected to change the main results. 

Figure \ref{fig:dustmodels} shows the comparison of the dust image in EIR with both models. Both models are capable of reproducing the dust ring morphology seen in the data, with normalized dips as low as 40\% of the maximum. 

\begin{figure}[!ht]
\begin{center}
\includegraphics[width=0.4\textwidth,trim=20 0 20 0]{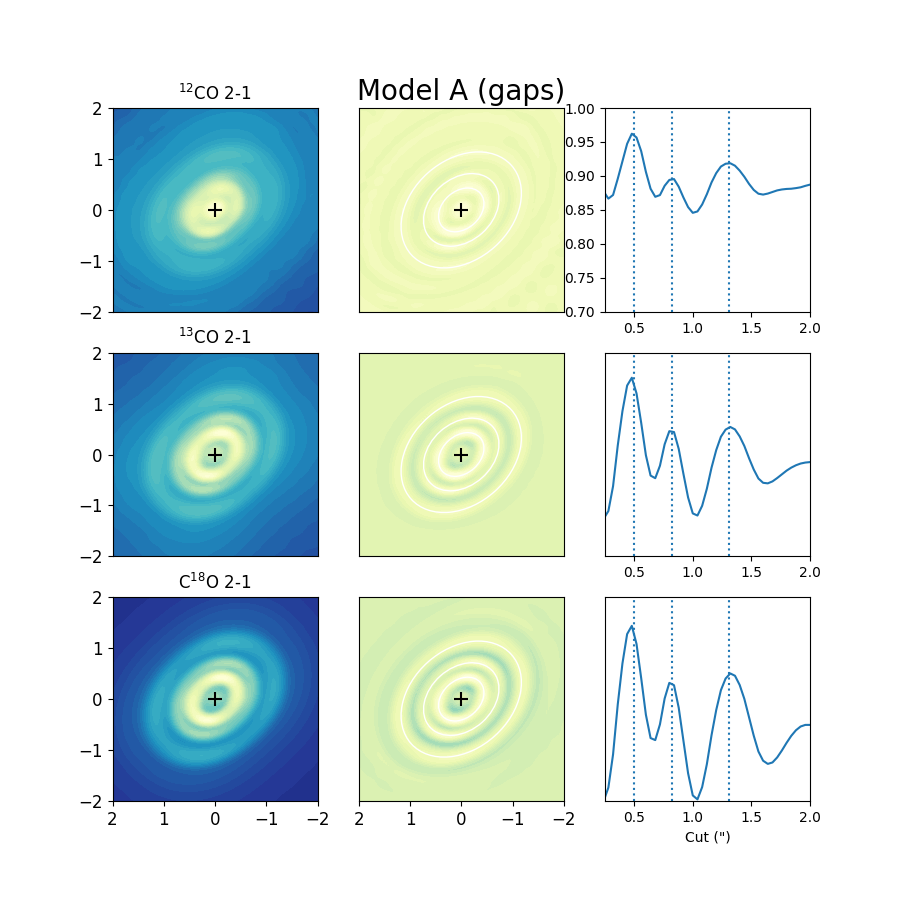}
\includegraphics[width=0.4\textwidth,trim=20 0 20 0]{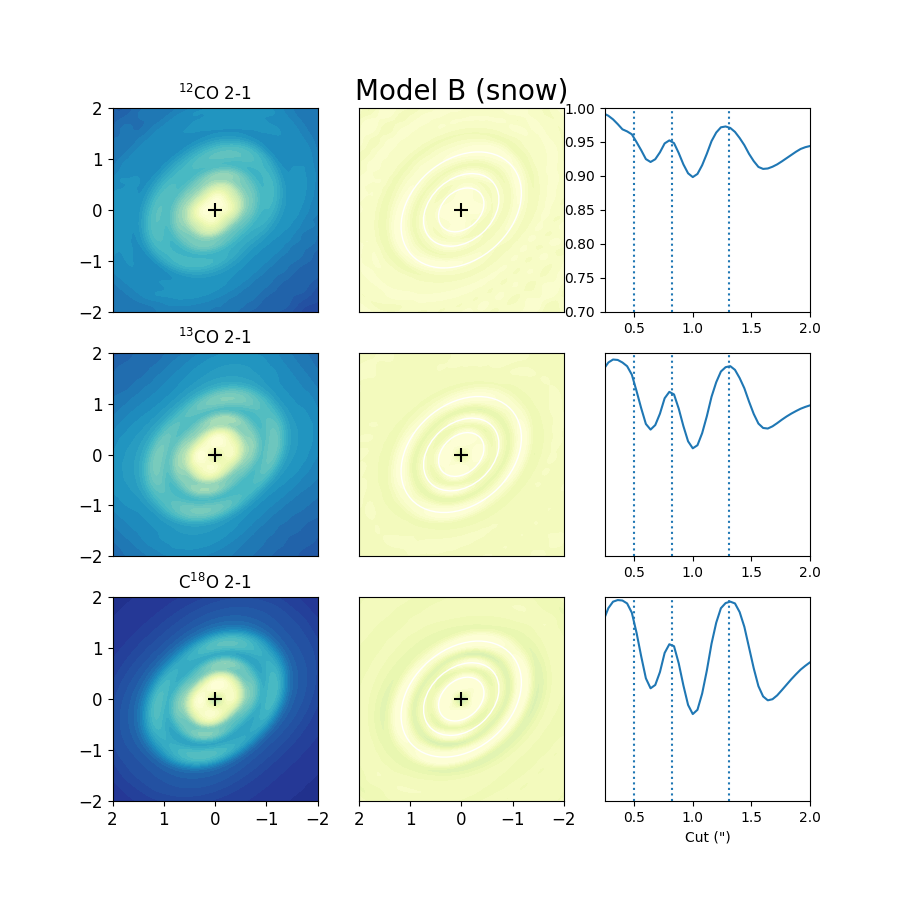}
\includegraphics[width=0.12\textwidth,trim=20 20 20 0]{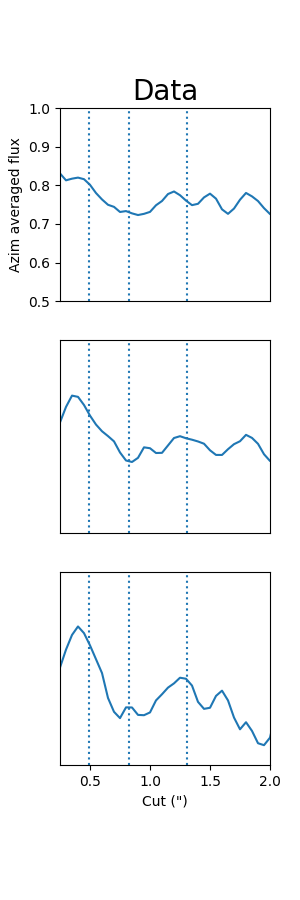}
\caption{CO emission as predicted by Model A (gap model, left) and Model B (snow model, right) and the CO emission in the data. From top to bottom the results for the integrated maps of $^{12}$CO 2--1, $^{13}$CO 2--1 and C$^{18}$O 2--1 are shown. The images have been convolved with the beam of the observations and shown in EIR. For each model, the first and second column show the model maps in original and EIR and the right column the azimuthal cut of the latter map. The dashed vertical lines and dashed ellipses indicate the center of each gap as defined in Table \ref{tbl:params}. This figure illustrates that the dust structure in the snow model does not affect the CO emission inside the gaps, while the gap model does.}
\label{fig:modelsCO}
\end{center}
\end{figure}

In Figure \ref{fig:modelsCO} the resulting CO emission for both models is shown for all three CO isotopologues. The inner part ($<$0.25") can be ignored as this is dominated by the continuum oversubtraction. The modeling does not try to account for this effect with an exact fit to the continuum, and is focused on the rings, not the center. The images are shown in EIR. In the azimuthally averaged cuts it immediately becomes apparent that the snow model results in marginal radial changes, only due to the continuum oversubtraction, while the gap model shows strong incremented rings in the CO. Remarkably, even though the gas surface density has been \emph{decreased} inside the dust gaps, the CO emission (in particular the C$^{18}$O) is \emph{brighter}. In order to understand this phenomenon, a more detailed look into the physics and chemistry in the disk is required. 

In Figure \ref{fig:DALIplots} the structure of each model as calculated by DALI is shown. Both models are compared with a model of similar disk mass where no gaps in dust or gas are introduced (Model 0).

\begin{figure}
\includegraphics[width=\textwidth]{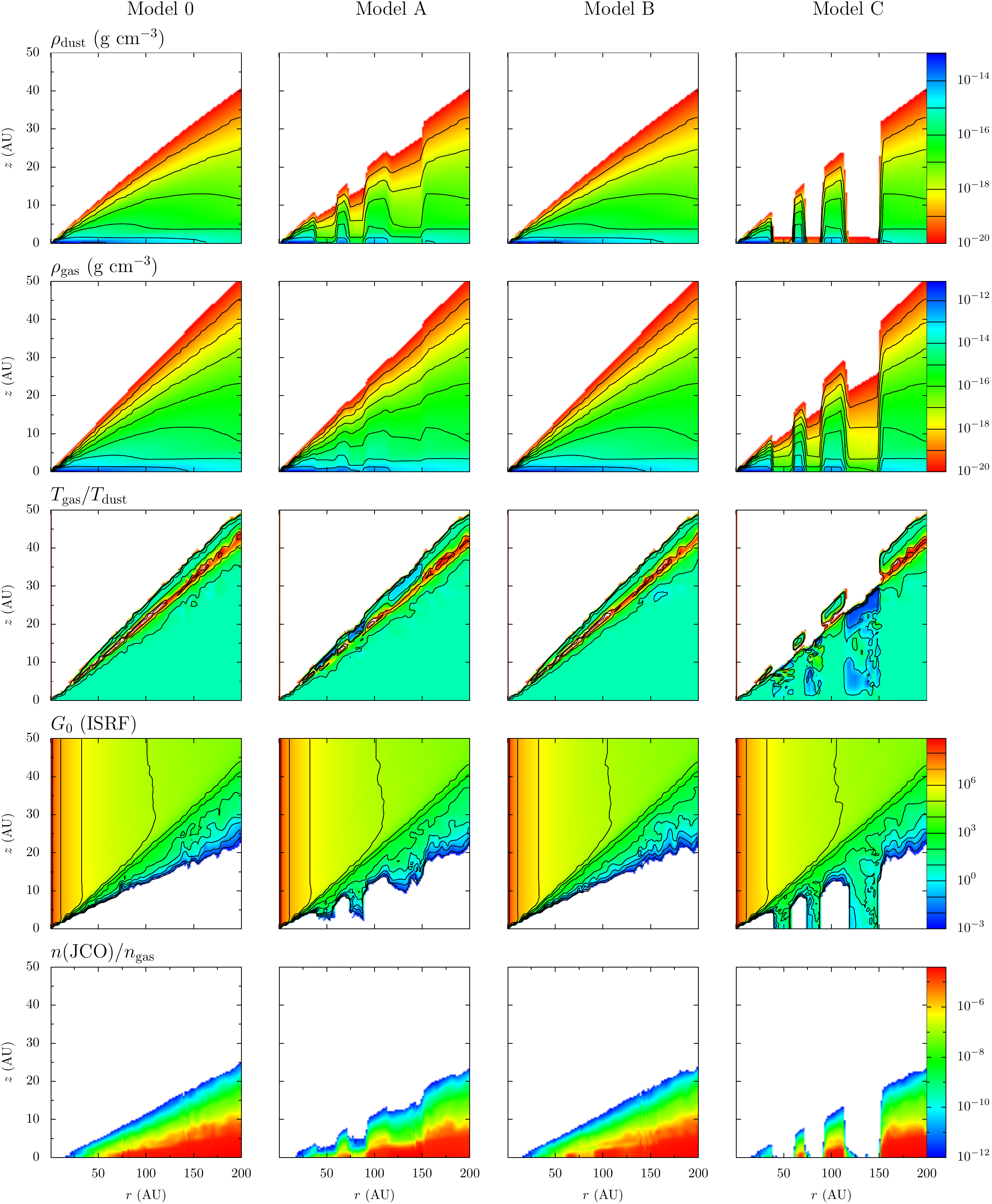}
\caption{DALI structure of each model. Model 0 is a model without any gaps in dust or gas; Model A is the gap model; Model B is the snowline model; Model C is like Model A but with much deeper gaps. From top to bottom the gas density, dust density, $T_{\rm gas}/T_{\rm dust}$ ratio, UV-field ($G_0$) and CO-ice abundance (JCO) is given.}
\label{fig:DALIplots}
\end{figure}

Comparing the models in Figure \ref{fig:DALIplots} reveals several remarkable differences in the temperature structure when gaps are introduced. Both $T_{\rm gas}$ and $T_{\rm dust}$ increase inside the gaps due to an increase in the UV field ($G_0$). The ratio $T_{\rm gas}/T_{\rm dust}$ slightly decreases in the surface layers, but not in the bulk region below the surface. We note that the temperature inversion in the thin upper surface layer (above the hot layer) in the $T_{\rm gas}$ is not relevant for the molecular gas temperature, as molecules such as CO and H$_2$ are photodissociated here. 

The increase in $T_{\rm gas}$ inside the gaps results in a decrease of frozen out CO molecules: in Model A, CO is abundant all the way down to the midplane in the gaps, whereas Model 0 shows that CO starts to be frozen out from 60 AU onward if no gaps are introduced. This leads to an increase in CO abundance inside the gaps, which produce the bright CO rings seen in Figure \ref{fig:modelsCO}. The CO emission is increased more for the less abundant C$^{18}$O than for $^{12}$CO, consistent with an abundance increase. This can only be explained by a marginal change in gas temperature in combination with a large change in abundance, as even the C$^{18}$O emission is marginally optically thick. Isotope-selective photodissociation \citep{Miotello2014} does not significantly change the outcome due to the high CO abundances. 

Results of hydrodynamic simulations of planet-disk interaction show an additional increase of the gas temperature up to 50\% in the gap as the result of the planet presence, as hydrostatic equilibrium needs to be maintained at a gap edge \citep[][Figure 6]{Isella2018}, which is further strengthening this case. As a combination of physical-chemical modeling to the level of DALI together with hydrodynamical simulations such as in \citet{Isella2018} is computationally unfeasible, it is not possible to calculate the combined effect quantitatively.

Model B on the other hand does not show any changes in temperature, UV or CO abundance structure despite the radial changes in the dust grain size distribution, which is consistent with what is seen in the CO images in Figure \ref{fig:modelsCO}. The snow model can thus reproduce dust rings without changing the CO emission itself.

\section{Discussion and conclusions}

Both the gap model and snow model can reproduce the morphology of the continuum image, reducing the emission inside the gaps similar to the data. However, the CO emission can distinguish between the two models: a snow model does not show any significant radial changes in the CO emission whereas a gap model can result in \emph{increased} CO emission inside the gap due to increased gas temperature, depending on the gap depth. The image data does not show an increase of emission inside the gaps, so the estimated density gap depths by \citet{Isella2016} are likely underestimated. 

In order to investigate how deep the gas gaps need to be in order to decrease the CO emission inside the gaps, a number of models with different $\delta_{\rm gas}$ are run and presented in Figure \ref{fig:deltagas}. The values of $\delta_{\rm gas}$ range between 1 and 10$^{-2}$. The deeper gap appears to be more consistent with the data. The gap depth in gas generally scales with the planet mass \citep{Fung2014,Kanagawa2015}: a deeper gap results in a larger planet masses. \citet{Isella2016} estimated Saturn-like planet masses of 0.1, 0.3 and 0.3 $M_{\rm Jup}$ respectively based on their gap depths, but our models suggest that the gaps, if caused by planets, must contain planets which are at least a Jupiter mass. This is also consistent with the findings of the kinematics \citep{Teague2018}. Furthermore, the larger planet masses are more consistent with the observed gap widths, which are $\sim$25 AU. Planets of $\sim$0.1 $M_{\rm Jup}$ are expected to open gaps of only a few AU wide \citep[e.g.][]{Dong2017gaps}, depending on the $\alpha$ viscosity. \citet{Isella2016} proposed multiple planets to be responsible for the wider gaps. However, if the dust is trapped in pressure maxima, the width of the dust gaps may be overestimating the actual gaps opened by the planet.

An observational distinction between the two models remains challenging: the density drops change the radial CO emission depending on the depth, but if the emission shows only no significant radial variations, both a shallow gap and snow model remain possible. However, it is also still possible that a single planet is responsible for multiple gaps, if the viscosity of the disk is very low \citep{Dong2017multi}, but there is no observational difference in the gas surface density in this case.

Another interesting question is what happens to the gas temperature in very deep gaps. \citet{Facchini2017gaps} have shown that a gap created by a Jupiter-mass planet creates a gap almost completed by dust, increasing the gas-to-dust ratio inside the gap, which results in $T_{\rm gas}<T_{\rm dust}$ in the entire gap due to the thermal decoupling between gas and dust. In the most right panel of Figure \ref{fig:DALIplots} we explore this scenario, with a drop in dust density of 10$^{-5}$ and gas density of 10$^{-3}$ inside the gaps. The temperature inversion inside the gaps is reproduced, but as there is so little dust in the gaps, there is no increase of frozen out CO molecules in this case. The low gas surface density (and resulting low CO abundance) result in deep CO gaps. 

\begin{figure}[!ht]
\begin{center}
\includegraphics[width=0.6\textwidth]{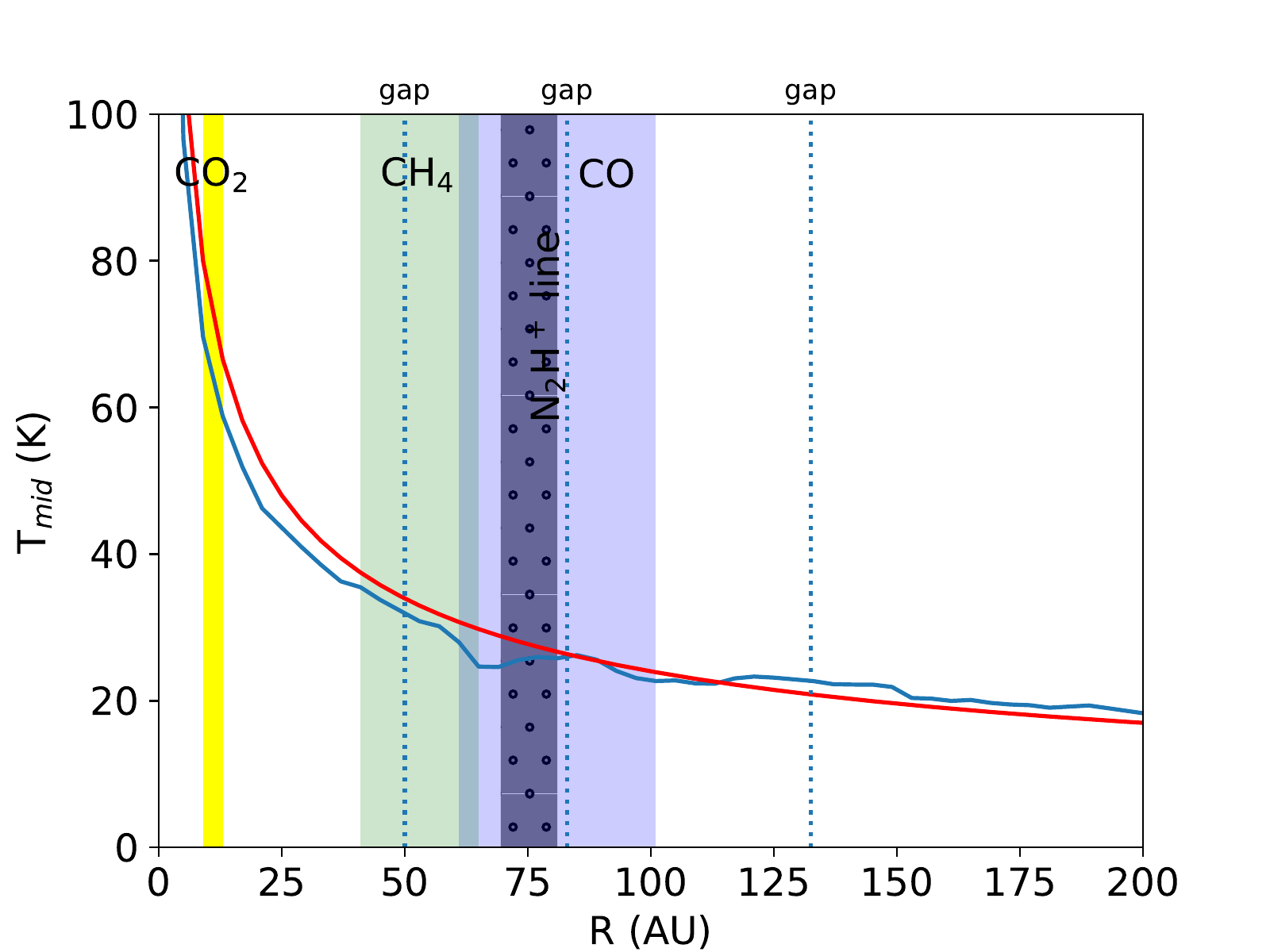}
\caption{Midplane temperature as function of radius of HD~163296 as computed in Model B (blue line). The red line is the parametrized midplane temperature as derived by \citet{Isella2016}. Dashed lines indicate the gap locations. Overlaid are the temperature regions where three main molecules in disks are expected to be frozen out: CO, CO$_2$ and CH$_4$. Furthermore, we overlay the snowline as derived from N$_2$H$^+$ and DCO$^+$ observations \citep{Qi2011}.}
\label{fig:snowlines}
\end{center}
\end{figure}

In summary, CO emission in dust gaps due to planets are complicated to interpret and one has to be careful with interpreting either increase or decrease of emission inside the gap. When no radial change is visible in the CO emission, this can be either a specific gas surface density drop, or an indicator of a snowline, as shown in Figure \ref{fig:modelsCO}. As the interpretation of the data of HD~1632926 is inconclusive about the presence of gaps in CO, the snowline scenario is still a possibility. This would also be more consistent with the observed presence of gaps and rings across the wide range of disk ages (0.4-10 Myr).

In order to explore the possibility of snowlines in HD~163296, we check the gas temperatures close to the midplane of Model B and compare them with condensation temperatures of molecules commonly found in disks, similar to the procedure in HL~Tau by \citet{Zhang2015}. 

Figure \ref{fig:snowlines} shows the midplane temperature as function of radius, with the gaps indicated. Also the parametrized midplane temperature used by \citet{Isella2016} is plotted, which is almost identical to the temperature calculated by our radiative transfer code. On top of this plot, we show the regions where CO, CO$_2$ and CH$_4$ are expected to be frozen out, based on the condensation temperatures derived in Table 2 in \citet{Zhang2015}: 23-28 K for CO, 26-32 K for CH$_4$ and 60-72 K for CO$_2$ respectively. Furthermore, we overlay the snowline as derived from N$_2$H$^+$ and DCO$^+$ observations \citep{Qi2011} in HD~163296, located at 75$\pm$6 AU.

It is clear that the CH$_4$ freeze-out region overlaps with the first gap, and the CO freeze-out region with the second gap, even further motivated by the location of the snowline as derived from the N$_2$H$^+$ data. The CO$_2$ freeze-out region does not overlap with a dust gap, but at this radius the dust is very optically thick so radial variations would not be visible, or perhaps remain unresolved. No molecule is known to freeze out in the temperature range of the third gap (18-20 K) but it is possible that the estimates for the condensation temperature N$_2$ are slightly underestimated in \citet{Zhang2015} at 12-16 K and the region is actually coinciding with that condensation region. Gaps caused by freeze-out zones may appear to be controversial, but this implies that the increased stickiness at these radii results in growth of dust particles well \emph{beyond} millimeter sizes, hence creating dips in the millimeter continuum emission.In that case, rings are simply the region where millimeter grains still remain. This is the opposite of our Model B and the results of \citet{Pinilla2017-ice}, where the snowline is suggested to coincide with the ring location rather than the gap location. Growth up to decimal sizes to explain gaps in dust images was originally proposed by \citet{Zhang2015}. The uncertainties in parameter choices such as viscosity, gas surface density and time scales imply that growth to decimal sizes cannot be excluded.

For HD~163296, the origin of the dust gaps thus remains unclear. Deeper CO observations at high spatial resolution are required to see whether gaps can be ruled out. This study shows the complicated nature of CO emission inside gaps, and the need for full physical-chemical modeling to interpret the underlying gas density.

  \begin{acknowledgements}
  The authors would like to thank Andrea Isella and Greta Guidi for providing the reduced fits cubes of HD~163296.
  This research was supported by the Munich Institute for Astro- and Particle Physics (MIAPP) of the DFG cluster of excellence "Origin and Structure of the Universe. This paper makes use of the
  following ALMA data: ADS/
JAO.ALMA\#2013.1.00601.S. ALMA is a partnership of ESO (representing its member states), NSF (USA) and
  NINS (Japan), together with NRC (Canada) and NSC and ASIAA (Taiwan),
  in cooperation with the Republic of Chile. The Joint ALMA
  Observatory is operated by ESO, AUI/NRAO and NAOJ.  \
  \end{acknowledgements}

\bibliographystyle{aasjournal}


\appendix
\section{Model properties}

\begin{table}[!ht]
\begin{center}
\caption{Model parameters}
\label{tbl:params}
\begin{tabular}{llll}
\hline
Parameter&Value&Description\\
\hline
\multirow{4}{*}{Stellar}&$L_*$&20.8 L$_{\odot}$&Stellar luminosity\\
&$T_{\rm eff}$&9250 K&Stellar temperature\\
&$\dot{M}_{\rm acc}$&10$^{-6.3}$ M$_{\odot}$ yr$^{-1}$&Accretion rate\\
&$d$&122 pc&Distance\\
\hline
\multirow{8}{*}{Disk}&$M_{\rm gas}$&3.9$\times$10$^{-1}$ M$_{\odot}$&Disk gas mass\\
&$M_{\rm dust}$&3.9$\times$10$^{-3}$ M$_{\odot}$&Disk dust mass\\
&$r_c$&150 AU&Critical radius\\
&$\Sigma_c$&25 g cm$^{-2}$&Surface density at $r_c$\\
&$h_c$&0.05&Scale height\\
&$\psi$&0.1&Flaring angle\\
&$r_{\rm out}$&450 AU&Outer radius\\
\hline
\multirow{3}{*}{Gaps}&$r_1$&40--60 AU&Dust gap 1\\
&$r_2$&74--92 AU&Dust gap 2\\
&$r_3$&114--151 AU&Dust gap 3\\
\hline
\end{tabular}
\end{center}
\end{table}



\begin{table}[!ht]
\begin{center}
\caption{Observational properties}
\label{tbl:obs}
\begin{tabular}{lllll}
\hline
Component&$F_{\rm tot,data}$&Beam size&$F_{\rm tot,snowmodel}$&$F_{\rm tot,gapmodel}$\\
&(Jy km s$^{-1}$)&(")&(Jy km s$^{-1}$)&(Jy km s$^{-1}$)\\
\hline
Continuum (230 GHz)&0.732&0.25$\times$0.25&0.95&1.27\\
$^{12}$CO 2--1&41.05&0.22$\times$0.16&25.50&25.65\\
$^{13}$CO 2--1&15.53&0.23$\times$0.17&10.47&10.78\\
C$^{18}$O 2--1&5.46&0.24$\times$0.17&4.465&4.471\\
\hline
\end{tabular}
\end{center}
\end{table}

\begin{figure}[!ht]
\begin{center}
\includegraphics[width=\textwidth]{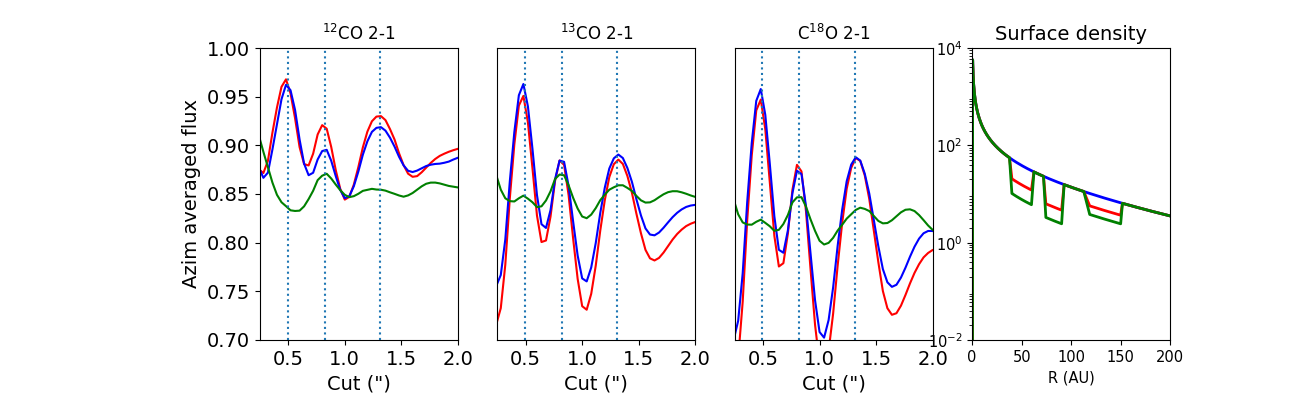}
\caption{CO emission as predicted by our gap model for different values of $\delta_{\rm gas}$ . The left three panels show the azimuthal averaged cuts of the EIR maps for the three CO isotopologues, convolved with the data resolution. The colors of the plot correspond to the models with different $\delta_{\rm gas}$ values and the density profiles as shown in the most right panel. The dashed lines indicate the locations of the gaps. A marginal drop in gas density consistent with Saturn-like planets results in an increase in CO emission, whereas a deeper drop (consistent with Jupiter-like planets) will result in a drop in CO emission.}
\label{fig:deltagas}
\end{center}
\end{figure}

\end{document}